\documentclass[useAMS,usenatbib,usegraphicx]{mn2e}
\usepackage{times}
\usepackage[]{natbib}
\usepackage{amssymb}

\title[Gravitational and mass distribution effects II]{Gravitational and mass distribution effects on stationary   superwinds II. Extended dark matter haloes}
\author[G.A. A\~norve-Zeferino, M.G. Corona-Galindo]{G.A. A\~norve-Zeferino$^{}$\thanks{E-mail:
ganiorve@inaoep.mx}, M.G. Corona-Galindo\\
Instituto Nacional de Astrof\'isica, \'Optica y Electr\'onica (INAOE),  Apdo. Postal 51 y 216,  72000, Puebla, Pue.,  M\'exico \\ }
\begin{document}

\date{Accepted date. Received date; in original form  date}

\pagerange{\pageref{firstpage}--\pageref{lastpage}} \pubyear{2010}

\maketitle

\label{firstpage} 

\begin{abstract}
In this second part, we generalize  the results of the previous paper. We    present an analytic superwind solution considering extended gravitationally-interacting dark-matter and baryonic  haloes. The  incorporation of the latter is critical, since they can   have a substantial effect  on  the hydrodynamics of superwinds generated by  massive galaxies. Although the presence of   extended and massive  haloes does not change the limit for   the   closed-box  enrichment of galaxies established in the first paper, they can trigger  an earlier activation of the open-box enrichment scenario, since their gravitational potentials can contribute  to the inhibition of the   free superwind. Moreover, the incorporation  of the extended haloes  will also enhance the  physical  setting behind   the superwind model, as we consider mass distributions with properties   that  emulate the results of recent   simulations of  $\Lambda$CDM haloes.

 \end{abstract}


\section{Introduction} \label{Intro}

In the first part of this series  of papers (A\~norve-Zeferino \& Corona-Galindo 2010; hereafter Paper I), we presented  a simple, spherically-symmetric  galactic  superwind  model considering  non-uniform dynamical mass distributions with analogue energy  and mass injection rates.  Galaxies were modeled in terms of four parameters: a characteristic  object radius,  $r_{\rm sc}$; the effective energy deposition rate, $\dot E_{\rm eff}$;  the effective mass deposition rate,  $\dot M_{\rm eff}$; and a normalized spatial distribution, $\rho_{\rm  s}$. The latter defined the distributions of the dynamical  mass  and of the mass and energy injection rates within $r<r_{\rm sc}$.

The  spatial distribution was assumed to follow a  truncated version of the Dehnen profile (1993), which  allows to recover  truncated versions of a plateau-like and  the Hernquist  (1990) and Jaffe (1983) profiles  as particular cases. As an initial simplification,  we considered only  the dynamical mass contained within $r_{\rm sc}$, which was assumed to account for most of the galaxy  dark matter (DM)  and baryonic mass (BM). This  simplification   allowed us  to make a direct comparison between our  analytic formulation  and  the numerical results  of Silich et al. (2010), whom assumed  a uniform distribution of the relevant galaxy parameters. Since in their model  they also considered  only the  gravitational field of the central region, we were able to analytically reproduce their numerical results as a particular case.  However,  the previous  assumption is a maximal extrapolation of  the results of Persic, Salucci \& Stel (1996) and Salucci \& Persic (1997), whom found that  in \emph{some}  cases the presence of  dark matter begins to be important well within the galaxy optical radius, $R_{\rm opt}$.  They estimated that from the galaxy centre{{s}} up to $R_{\rm opt}$, the fraction of DM goes from 0\%  up to 30\%-70\% (see also Salucci et al. 2007). 

Thus,  the initial supposition, as well as  the assumption of spherical symmetry, can only be adequately interpreted  in the context of  a zeroth-order  approximation for evaluating 
the effect of the gravitational field of ellipsoidal galaxies on  the inner  superwind solution ($r\le r_{\rm sc}$).  Similarly,  the  zeroth-order distortion  of the external superwind hydrodynamical profiles ($r>r_{\rm sc}$) is only adequately predicted for  galaxies with either   low masses  or  very diffuse  DM  and BM haloes.

For disc galaxies with large masses, the aforementioned simplification certainly does not hold, as the  observed flat rotation curves of  the extended discs \emph{require} a significant amount of  dark and baryonic matter outside of the bulge ($ r_{\rm sc}\sim r_{\rm bulge}$).  Needless to say,  the presence of  discs will also produce collimated flows. Furthermore, extended and fairly  massive  haloes   can also have important repercussions on the superwind solution for the case of  ellipsoidal galaxies ($r_{\rm sc}\sim r_{\rm nucleus}$),  according to the    properties of DM haloes derived by Persic et al. (1996) and  Salucci \& Persic (1997). 

Nevertheless, the zeroth-order   superwind  solution that we presented in Paper I has the advantage of being analytic. Thus, it can be used to construct the solution for the case in which departures form spherical symmetry are important. In order to do this adequately for a wide range of galaxy masses, we need first to incorporate the effect of the external DM and BM halo under the assumption of spherical symmetry.

In this work, we obtain such a solution considering gravitationally-interacting external haloes (Section \ref{shalo}). We consider halo distributions that emulate the results of recent  cosmological simulations (Section \ref{halo}). Later, in Section \ref{thres}, we obtain new limits for the open-box enrichment scenario and the existence of   accelerating superwind solutions (see  Paper I). Finally, in Section \ref{hydro}, we evaluate the impact of the halo gravitational potential on the superwind hydrodynamical profiles. The conclusions are presented in Section \ref{con}.

\section{An analytic solution for the free superwind including  extended haloes} \label{shalo}

When the   haloes are included, the equation of conservation of energy outside of the galaxy characteristic radius, equation (9) in Paper I,  transforms into 

\begin{equation}
\frac{1}{r^2}\frac{{\rm d} \left[\rho u r^2\left(\frac{1}{2}u^2+(\eta +1)\frac{P}{\rho}\right) \right] }{{\rm d}r}= -\rho u ( \nabla \phi + \nabla \phi_{\rm h}). \label{CE}
\end{equation}

Above, the hydrodynamical variables are represented by their usual symbols, $\eta$ is the polytropic index   and  $-\nabla\phi=-GM_{\rm DM}/r^2$,  where  $M_{\rm DM}$  is the total dynamical mass   within $r_{\rm sc}$. Similarly, $-\nabla \phi_{\rm h}=-GM_{\rm h}(r)/r^2$, where $M_{\rm h}(r)$ is the cumulative dynamical mass (i.e. DM+BM) of the external halo, which  has a total mass $M_{\rm H}$.  We will allow the  profile of the external  halo to be defined  either as a continuation of the internal profile or as  a  centrally truncated  profile with different characteristics. 

The integration of equation (\ref{CE}) yields a Bernoulli-like equation

\begin{equation}
\frac{1}{2} u^2 + (\eta+1) \frac{P}{\rho}  = \frac{1}{2} V_{\rm g}^2 - \tau \phi - \phi_{\rm h},\label{CEi} 
\end{equation}

\noindent where $\phi=-GM_{\rm DM}/r $,  $\phi_{\rm h}(r)$ is the gravitational potential at $r>r_{\rm sc}$ associated to the non-truncated version of the external halo, $\tau=1-M_{\rm H}/M_{\rm DM}$ accounts for  truncation effects, and   $V_{\rm g}$  is the asymptotic terminal speed, which is  given by 
\begin{equation}
\frac{1}{2} V_{\rm g}^2 = \frac{1}{2} V_{\infty}^2  -\frac{\tau GM_{\rm DM}}{r_{\rm sc}}+  \phi_{\rm h}(r_{\rm sc})-\frac{\left(1+\frac{1}{A}\right)}{(5-2\alpha)}\frac{GM_{\rm DM}}{r_{\rm sc}}.\label{CEii}
\end{equation}

In the last equation, $V_\infty$ is the effective terminal speed due to the thermalization of SNe ejecta and individual stellar winds inside of the galaxy, $\alpha$ determines the steepness of the central dynamical mass distribution ($r<r_{\rm sc}$)  and $A$ its concentration [see equation (11) in Paper I].  Note that  $\tau=0$ implies an uninterrupted,  \emph{continuous} gravitational potential.  Similarly, $\tau<0$ implies a centrally truncated external halo with a mass larger than $M_{\rm DM}$, and $0<\tau<1 $ implies the opposite. When $\tau=1$ there is no external halo, and thus $\phi_{\rm h}$ is identically zero.

 As in Paper I, we will  work in terms of   dimensionless variables. For the present case they are:

\begin{equation}
R=\frac{r}{r_{\rm sc}}, \; U= \frac{u^2}{V_{\rm g} ^2}, \;   \mbox{and}\;  \Phi=-\tau\frac{V_{\rm eg}}{R} + \Phi_{\rm h} (R);
 \end{equation}

\noindent where $V_{\rm eg}$ is given by

\begin{equation}
V_{\rm eg}= \frac{v_{\rm e}^2}{V_{ \rm g}^2}=\frac{2GM_{\rm DM}}{r_{sc}V_{\rm g}^2},
\end{equation}

\noindent  and $\Phi_{\rm h}(R)$ is $\phi_{\rm h}(r)$ written in terms of $R$ and normalized to $V_{\rm g}^2/2$. The conservation laws can now be reduced to the \emph{same} governing differential equation than in Paper I, see its equation (43).

Thus, within  the theoretical framework developed in Paper I, it is very easy to prove  that the  transonic free superwind solution  is given by

 \begin{equation}
R= D U^{-1/4} \left[ 1-U +\tau\frac{V_{\rm eg}}{R} -\Phi_{\rm h} (R)\right]^{-\eta/2}, \label{gensol}
\end{equation}

\noindent with

\begin{equation}
D=\left(\frac{1}{2\eta+1}\right)^{1/4} \left\{ \frac{2\eta\left[1+\tau V_{\rm eg} -\Phi_{\rm h}(1)\right]}{2\eta+1} \right\}^{\eta/2}.
\end{equation}

Again, as in our previous work,  we will  give preference to the parametric version of the solution:

\begin{equation}
x[x-\tau V_{\rm eg}+R\Phi_{\rm h}(R)]^{\frac{2\eta-3}{4}}=  D_0(y+1)^{-\frac{1}{2\eta+1}}(2\eta-y)^{-\frac{2\eta}{2\eta+1}}, \label{x-y-eta}
\end{equation}

\noindent where $y$ is a parameter that varies between 0 and $2\eta$ and

\begin{equation}
x= R+\tau V_{\rm eg} \label{x} - R\Phi_{\rm h}(R), \label{x}
\end{equation}

\begin{equation}
y=\left[\frac{(2\eta+1)U}{1+\tau \frac{V_{\rm eg}}{R} -\Phi_{\rm h}(R)}-1\right], \label{y}
\end{equation}

\noindent and

\begin{equation}
D_0=  [1+\tau V_{\rm eg}-\Phi_{\rm h}(1)](2\eta)^{\frac{2\eta}{2\eta+1}}. \label{D0}
\end{equation}

To obtain the hydrodynamical profiles, one just need{{s}} to follow the algorithm presented  at the end of section 3.3 in Paper I. An advantage of the parametric solution is that it allows to work with just functions of $R$ in the first two critical steps,  related to equations (\ref{x-y-eta}) and (\ref{x}). On the other hand, equation (\ref{gensol}) involves both $R$ and $U$. For $\eta=3/2$ (equivalent to the case   $\gamma=5/3$, where $\gamma$ is the adiabatic index) there is no need for a numerical root finder in the first step of our algorithm. In the second step however,  its use will be most likely unavoidable, as the particular form of the assumed  gravitational potential (i.e. of the external halo profile) is  involved. 

In Section \ref{thres}, we  will give the limit above which the stationary solution is disrupted in the external zone ($r>r_{\rm sc}$) and the necessary condition for  an accelerating stationary superwind solution. In order to do this,  we  will   specify  first the normalized potential $\Phi_{\rm h}$ in the next section.

\section{The extended halo profiles} \label{halo}

How are the DM  and BM  distributed\footnote{As in Paper I, we will assume that their distributions  are analogue.} outside of the galaxy characteristic radius? Since we have  permitted   centrally truncated profiles  for the external  halo, theoretically,  we can choose practically any of the usually assumed  distributions; e.g. a NFW profile,   Navarro, Frenk \& White 1997; a generalized NFW profile,  Moore et al. 1999;  a self-similar profile, Yoshikawa \& Suto 1999;  an isothermal profile, and so on. Given that      the most commonly used profiles depend  on at least two parameters, and given also the additional freedom introduced by our truncated {{halo}} scheme; there is a vast number  of  profiles and  parameters that  can give reasonable agreement with observational studies and with the predictions of cosmological simulations.

We will try to rely on physical insight for  selecting the external halo profile that we will  use in our model.  Recent  cosmological simulations carried out by Abadi et al. (2010) predict that dark matter haloes always contract as a result of galaxy formation. They also found that the contraction  effect is substantially  less pronounced than predicted by the adiabatic contraction model (Blumenthal et al. 1986). On similar grounds, according to  the high-resolution N-body cosmological simulations of  $\Lambda$CDM haloes carried out by  Navarro et al. (2010), the departures from similarity  in the velocity dispersion   and density  profiles   correlate in such a way,  that  a power law for the spherically averaged pseudo-phase-space density is preserved,  $\rho/\sigma^3\propto r^{-1.875}$. They remarked that the  index of the previous power law is identical to  that of a Bertschinger's similarity solution for self-similar infall onto a point mass (in an Einstein-de Sitter Universe). 
They conclude that $\Lambda$CDM  haloes are not strictly universal, but that  the  departure from similarity previously mentioned may be a  fundamental structural property.

Bearing  in mind the results described above,  we deduce that the cases $\tau<0$ and $0<\tau<1$ correspond to  artificial mathematically induced  constraints  that make  continuous the potential at $r=r_{\rm sc}$ for arbitrarily-chosen external-halo  profiles [see equation (\ref{CEi})]. The case $\tau=1$ is physical, but corresponds to the zeroth-order approximation regarded as inadequate for some galaxies in Section \ref{Intro}. Thus, the case $\tau=0$ is of special interest, as it implies an unforced continuity of the gravitational potential. It turns out that adequately chosen truncated Dehnen profiles satisfy naturally the latter condition.

For  $r<r_{\rm sc}$,  the cumulative dynamical mass  corresponding to a truncated Dehnen profile\footnote{See  also equation (3) in Dehnen (1993).} is given by equation (13) in Paper I: 

\begin{equation}
M(r) = M_{\rm DM}(1+A)^{3-\alpha}\left(\frac{R}{R+A}\right)^{3-\alpha}.
\end{equation}

We will also assume a truncated Dehnen profile for the external halo, but we will demand a cumulative mass of the form: 
\begin{equation}
M_{\rm h}(r) = M_{\rm DM}(1+A_1)^{3-\alpha_1}\left(\frac{R}{R+A_1}\right)^{3-\alpha_1}. \label{Moutt}
\end{equation}

 At $R=1$ we have that $M(1)=M_{\rm h}(1)$. Note that for this, we do not require  $A_1=A$ nor $\alpha_1=\alpha$. The last property  can be interpreted in terms  of a contraction of an initial spatial configuration of  DM and BM with concentration $A_1$ and steepness $\alpha_1$ which  produced  a new configuration with concentration $A$ and steepness $\alpha$ for  $r<r_{\rm sc}$, or well, vice-versa, if other processes {{were}} involved (v.gr. angular momentum).   On the other hand, a trivial but  important relationship can be obtained from the condition  $M(1)=M_{\rm h}(1)$ by separating the baryonic and dark matter components:

 \begin{equation}
r_{\rm sc} [M_{\rm bar} + M_{\rm dark}]=r_{\rm sc}  [M_{\rm bar} +M_{\rm dark}]_1.
 \end{equation}
 
\noindent This  could be interpreted as an integral equivalent of the equation for adiabatic collapse derived by Blumenthal et al. (1986). Additionally, given that  the radial velocity dispersion associated to the Dehnen profile goes as $\sigma\sim r^{\alpha/2}$ when $r\rightarrow 0$, we are able to recover the index of the  Bertschinger's  power law near the centre of the galaxy when  $\alpha=3/4$. However,  Navarro et al. (2010) obtained the index from radial averaging, which implies that $\alpha$ can adopt values within a wider range.

Note that  in turn, the previous  configurations could be interpreted as the result of the contraction of an unperturbed configuration   away from the galaxy. This is equivalent to saying that  a galaxy  formed from  the perturbation of an initial state ($A_0$,$\alpha_0$), and that  after certain time, the perturbation bifurcated and produced two inner contracted states characterized by  ($A$,$\alpha$) and ($A_1$,$\alpha_1$). The first state characterizes the inner regions of the  galaxy, $r<r_{\rm sc}$. Then, the characteristic radius $r_{\rm sc}$ can be taken either as   the radius of a     galaxy nucleus or of a bulge. The second state characterizes the outer portions of the galaxy  (e.g. a disc + DM). This is  in agreement with the aforementioned cosmological simulations,  and it implies that galaxies carved out gravitational potential  holes when they formed, and that they correspond to local depressions of  an otherwise smoother gravitational potential. 

Here, we are just interested in the superwind solution, so, in order to keep things simple,  we will just consider  the  states $(A,\alpha)$ and $(A_1,\alpha_1)$, i.e.  we will ignore the  depression  of the reference gravitational potential $(A_0,\alpha_0)$. The price that we will pay for this, as well as for the implicit analogue distribution of the baryonic and DM components assumed in our scheme, is that instead of (almost) 'perfectly' flat rotation curves  up to 15 times the optical radius (Persic et al. 1996, Salucci \&  Persic 1997), the rotation curves will show some downwards skewness at large radii. They {{are}} however well above the curves corresponding  to   keplerian rotation of the baryonic mass. Evenmore,   the  behaviour of the associated rotation curves  away from $r_{\rm sc}$ is consistent with that of the  universal  rotation curves derived by Salucci et al. (2007) for spiral galaxies. Anyway,  for our purposes, the behaviour at large radii is not that important, as the thermalization driven superwind solution is valid only close to the galaxy\footnote{This implies that the effect of   the 'real' $\Phi_{\rm h}$ can be emulated there by giving adequate values to $A_1$ and $\alpha_1$.}  (see e.g.  Strickland \& Heckman 2009). So, we will proceed to give the expression corresponding to the external gravitational potential.

By taking the limit $R\rightarrowÊ\infty$ in equation (\ref{Moutt}),   one finds that   the  total dynamical mass  is given by  $M_{\rm t}=M_{\rm DM} (1+A_1)^{3-\alpha_1}$.  The expression of the associated gravitational potential for $0\le \alpha\le1$ is then similar  to that given by equation (2) in Dehnen (1993):

\begin{equation}
\Phi_{\rm h}(R)= -\frac{V_{\rm eg}(1+A_1)^{3-\alpha_1}}{(2-\alpha_1) A_1}\left[1-\left(\frac{R}{R+A_1}\right)^{2-\alpha_1}\right].
\end{equation}

With this, we can establish new approximated  thresholds for the open-box enrichment scenario and for accelerating superwind solutions.

\section{Thresholds for open-box enrichment and accelerating superwind solutions} \label{thres}

When the effect of the external halo is considered, the asymptotic terminal speed is given by

\begin{equation}
V_{ \rm g} =\left[1 -\left(\frac{1+\frac{1}{A}}{5-2\alpha}\right)V_{\rm e} + \frac{V_{\rm g}^2}{V_{\infty}^2}\Phi_{\rm h}(1) \right]^{1/2} V_{\infty}. \label{Vg}
\end{equation}

\noindent The  flow enters into  non-stationary  regimes  (inpouring or outpouring, see Paper I)  when 

\begin{equation}
 \left(\frac{1+\frac{1}{A}}{5-2\alpha}\right)V_{\rm e}  - \frac{V_{\rm g}^2}{V_\infty^2}\Phi_{\rm h}(1) \ge 1. \label{outin}
\end{equation}

\noindent  When the above inequality holds,  the galaxy can eventually enter into an open-box enrichment scenario. Otherwise, we will have fully stationary solutions, unless  radiative cooling or self-gravitation inhibit the  stationary solution.

Fully stationary superwinds  have accelerating velocity profiles when

\begin{equation}
-(2\eta+1) \frac{V_{\rm g}^2}{V_\infty^2}\Phi_{\rm h}(1)+2\eta \left(\frac{1+\frac{1}{A}}{5-2\alpha}\right)V_{\rm e}\le 2\eta,
\end{equation}

\noindent otherwise, they have decelerating velocity profiles. When the equality holds in the above relation, we have an almost constant external velocity profile with characteristic velocity
 
\[V_{\rm g}= (2\eta)^{-1/2}[-2\phi_{\rm h}(r_{\rm sc})]^{1/2}= \]
\begin{equation}
\eta^{-1/2}v_{\rm rot}\left\{ \frac{ (1+A_1)^{3-\alpha_1}}{(2-\alpha_1) A_1}\left[1-\left(\frac{1}{1+A_1}\right)^{2-\alpha_1}\right]\right\}^{1/2},
\end{equation}

\noindent where $v_{\rm rot}$ is the rotation speed at $r_{\rm sc}$.

\section{Effect on the hydrodynamics} \label{hydro}

In our model, the  dynamical  mass ($M_{\rm DM}$)  contained within a bulge or galaxy nucleus experiencing an starburst episode  is related to the concentration parameter  and steepness of the external halo,   and to  the total dark matter and baryonic mass:

\begin{equation}
M_{\rm DM}= \frac{M_{\rm t}}{(1+A_1)^{3-\alpha_1}}.
\end{equation}

The dynamical mass contained in the external halo ($r>r_{\rm sc}$) is 

\begin{equation}
M_{\rm H}= M_{\rm t} \left[1 -\frac{1}{(1+A_1)^{3-\alpha_1}}\right].
\end{equation}

Similarly, the  dynamical mass contained up to an external characteristic radius (normalized to $r_{\rm sc}$), $R_{\rm D}$, is
\begin{equation}
M_{\rm D}=M_{\rm t}\left(\frac{R_{\rm D}}{R_{\rm D}+A_1}\right)^{3-\alpha_1}.  \label{Mout}
\end{equation}

The radius $R_{\rm D} $ can be associated to the  'disc' radius of the BM or  well  to the BM+DM virial radius. So, all the relevant galaxy parameters  are correlated, in a similar   fashion as in the  work of Salucci et al. (2007).  Nevertheless, we emphasize that the relationship between the parameters is alike but of course not the same, since here we constructed our theoretical model only  following the results of the simulations of Abadi et al. (2010) and Navarro et al. (2010). 

We will proceed to discuss the effect of the extended haloes on the hydrodynamics. In order to do this, we consider the hydrodynamical models  presented in Table \ref{Table1} and Table \ref{Table2}.  The first table gives the inner parameters for three galaxies with different characteristics. The second table gives the properties of their external haloes. The groundwork for the discussion will  be  the premise   that the spherical symmetric  superwind solution is a zeroth-order approximation to the aspherical case. We will consider again a reference effective terminal speed of 2500 km s$^{-1}$ for the case of  null mass-loading, fully efficient  thermalization, and total participation within the starburst volume,  i.e. for  $\epsilon=\beta=\zeta=1$.  For  models 2 and 3, the SFRs were obtained from formula (1) in Rupke, Veilleux \& Sanders (2005a) and formula (28) in Paper I, i.e. we  considered SFRs that are  consistent  with the  typical observed luminosities for the object types, and that  in  parameter space, place the objects below the threshold for catastrophic cooling. We find that the predicted temperature profile is barely modified by the presence of the extended haloes. However,  drastic changes are  produced  in the velocity profile.

The first model is an extended version of model 5 in Paper I,  and corresponds to a synthetic isolated dwarf elliptical  galaxy that tries to emulate the  characteristics of  the most massive outlier of the mass-metallicity relationship detected by Peeples, Pogge \& Stanek (2008, see also Paper I). We assumed that the galaxy formed by a contraction of    $\sim 40\%$ of an initially unperturbed   subhalo of DM and BM which had $\sim  70\%$ of its total mass   located within $r\sim 3r_{\rm sc}$, so we used  $A_1=0.5$, $\alpha_1=3/4$ and $R_{\rm D}=1$.  The latter is equivalent to saying that in this case there is no disc, i.e.  we only have a galaxy nucleus.  The internal dynamical mass distribution follows a plateau-like profile, which implies that  some mechanism --perhaps the action of early powerful superwinds associated to a more extended and powerful starburst episode (see Governato et al. 2010) or  internal dynamical processes--  has also transformed the initial mass configuration. In this model,  starburst activity still persists  near the galaxy centre, but with a high concentration.  We assumed a low thermalization efficiency, which implies a small number of massive stars and SNe within the characteristic concentration radius, $A=0.1$. The justification for  this is that the SFR is low,  and that although small, the concentration radius is much  larger than the typical radius of a massive star, i.e. the filling factor is low. Similarly, because of  the small number of massive stars, just a small incorporation of mass is necessary to produce a heavily  mass-loaded superwind. In this model, the presence of the extended halo suppresses the free superwind solution and the galaxy experiences an open-box enrichment [see equation (\ref{outin})] by keeping the metals processed by the few massive stars  still present near the galaxy centre. This will require however an already gas-poor galaxy  at the moment  at which the pollution occurred       (Peeples, Pogge \& Stanek, 2008).  As suggested above, the required low mass fraction could have been produced by  the action of early  superwinds  associated to   previous and more powerful starburst activity. This is consistent with the views of Peeples et al. (2008), which regarded their sample of outliers as transitional galaxies  in their way to  becoming typically isolated dE and dSph galaxies, but with a high metallicity.  The suppression of the free superwind solution is practically insensitive to {{the}} value of $0\le\alpha_1\le 1$, which indicates that the enrichment is produced by the physical conditions within $A$ and the  initial concentration of the unperturbed subhalo from which the galaxy formed.

\begin{figure}
\includegraphics[height=50mm]{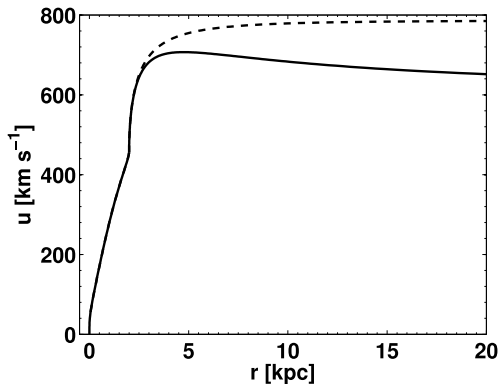}\caption{Superwind velocity profile for model 2 (solid line).  The dashed-line   represents the profile that would result if the external halo were neglected.  } \label{fig1}
\end{figure}

The second model considers the synthetic and very  massive blue compact dwarf  galaxy  modeled in  Paper I. Here, we add an extended disc to the model in order to 'transform' the galaxy into a  luminous infrared one\footnote{N.B. As LIRGs and ULIRGs,  BCDs may be the result of   mergers, although generally they have lower masses, given that   they mostly form  from the merging  of  dwarf galaxies. Nevertheless, on a higher end, luminous blue compact galaxies  can have dynamical masses of up to $\sim 10^{12}$ M$_\odot$ (Garland et al. 2004, Pisano et al. 2010).} (LIRG, $L_{IR} \sim 10^{11}$ L$_{\odot}$).  LIRGs and ULIRGs may be the end  result of  the merging   of two moderate-size spiral galaxies and display  traces of convergence to an  elliptical  morphology  (Sanders \& Mirabel 1996; Rupke et al.  2005a).  We will model  a LIRG that still  exhibit evidence of  heavily warped  and thick discs, displaying a morphology perhaps similar to that  of  the central component of Arp 299 (Sargent \& Scoville 1991; Heckman et al. 1999; Hibbard \& Yun, 1999;  Hu et al. 2004), but with just one nucleus. We assume that such features extend to up to 5 times the radius of the merger nucleus; thus, $R_{\rm D}=5$. The assumed mass and extension are consistent with  CO  emission observations of  (U)LIRGs (Lonsdale, Farrah \& Smith 2006 and references therein). We further assume that the merging process has similarly  transformed the steepnesses of the internal and external mass profiles of the  interacting galaxies unperturbed  haloes,  such that $\alpha=\alpha_1=3/4$.  We adopt the value $A_1=1$ since it produces interesting proportions. In such a case,  $\sim 66\%$ of the BM and DM  of both galaxies is  contained within the warped discs characteristic radius and about $\sim 30\%$ of this fraction resides within the merger nucleus (that is $\sim 20\%$ of the total mass).  As in the original model, starburst activity is present in the nucleus with a somewhat high concentration ($A=0.4$), the thermalization efficiency is 0.5 and mass loading is important, $\beta=3$ (see Heckman et al. 1999). In this model, the gravitational field of the  external halo transforms the accelerating superwind solution associated to the original model into a bounded decelerating one (Fig. \ref{fig1}). This effect occurs because now we have a more massive galaxy. The produced deceleration will enhance the observable properties of the superwind because of a proportional density increment.  However, an even larger total mass   could  inhibit the superwind solution. As mentioned in Paper I, this is consistent with the superwind  scaling properties found by Rupke, Veilleux \& Sanders  (2005b), whom reported and initial  increment of the superwind observable properties with galaxy mass  and a posterior flattening with the same.

 Rupke et al. (2005{{b}}) also reported a flattening of the superwind observable properties at high SFR.  In principle, the normalized free superwind solution (Section \ref{shalo}) is insensitive to the SFR (provided that {{it}} could be considered constant during a relatively large time interval), as it just depends on the effective  and asymptotic terminal speeds. However, high SFRs will  intensify the effect of radiative cooling, as more mass will be injected per unit  time and volume, and thus, the stationary solution could also be radiatively inhibited. We have  properly  addressed this issue {{in}} section 2.1.2 of Paper I.

\begin{figure}
\includegraphics[height=50mm]{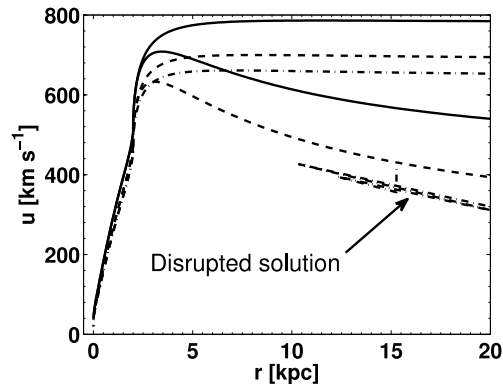}\caption{Velocity profiles for model 3. The solid lines corresponds to the parameters showed in Table \ref{Table1}.  For this parameters,   $V_{\infty}\approx 1208$   km s$^{-1}$. The lower (upper) solid line  (does not) consider(s) the presence of the external halo.  Similarly, the dashed and dash-dotted lines  correspond to $V_{\infty}=1150$ km s$^{-1}$ and  $V_{\infty}=1125$ km s$^{-1}$, respectively. For the latter case, the stationary free superwind solution does not exist.} \label{fig2}
\end{figure}
As an extreme example of the effect of the nominal value of the galaxy mass, we model a  massive and  'rare' radio galaxy with a very extended halo (see  e.g. Genzel et al. 2003). We consider a  galaxy with a dynamical  mass of  $1\times 10^{11}$ M$_\odot$ within its nucleus of  $r_{\rm sc}\sim 2$ kpc. A mildly concentrated starburst ($A=0.5$) is present in the nucleus, which has a cuspy dynamical mass distribution ($\alpha=1$).  We consider that the steepnesses of the inner region and the halo are the same and that the total mass of the galaxy is $M_{\rm t}=4\times 10^{11}$ M$_\odot$.  This requires that $A_1=1$. This implies that the half-mass radius is  $r\sim 2.5 r_{\rm sc}$ and that $\sim80 \%$ of the total mass is contained within $r\sim 8r_{\rm sc}$. In this model a high deceleration of the superwind is produced, and    the flow is unstable to small variations of the effective terminal speed (thermalization efficiency), as shown in Fig \ref{fig2}.  As a consequence, the flow could eventually enter into  the outpouring or even the inpouring regime. On the other hand, if instead of a continuous steepness, we consider that the typical  cuspy halo profile (with slope $\alpha=1$) resulted from the contraction of a smoother one, say with $\alpha_1=3/4$ and $A_1=1$, the  free superwind solution would be inhibited and the galaxy  could enrich itself with is produced metals in an open-box scenario. This would occur because in the second case,  the total mass is slightly larger, $M_{\rm t}=4.75\times 10^{11}$ M$_\odot$. The cumulative dynamical masses of   the two assumed external profiles are very similar, their ratio varies from a value of 1 at $r_{\rm sc}$  (they are identical as they must), up to a value $\sim 0.86$ at $r=10r_{\rm sc}$; nevertheless, such a small variation is enough to suppress the stationary superwind solution. This reflects the fact that at the limit of large galaxy masses, galaxies will retain most of their metals, as expected. 

 \begin{table*} 
\begin{minipage}{175mm}
\caption{Reference hydrodynamical models. Galaxy parameters for $r<r_{\rm sc}$.}
\begin{tabular}{@{}llllllllllllll}
\hline
Model  &   Type&$\alpha$  &$A$ & $r_{\rm sc}$ & $M_{\rm DM}$ &SFR&$\beta$& $\epsilon$ &  $\zeta$&$V_{\infty}$&  $V_{\rm e}$ &Regime (No halo)\\
&&&&(kpc)&($\times 10^{8}$ M$_{\odot}$)&  M$_{\odot}$ yr$^{-1}$& &&& km s$^{-1}$\\
&&$(\rm a)$&$(\rm b)$&$(\rm c)$&$(\rm d)$&$(\rm e)$&$(\rm f)$&$(\rm g)$&$(\rm h)$&$(\rm i)$&$(\rm j)$&$(\rm k)$\\
 \hline
                   1&  dE& 0 & 0.1        &1  &  100   &   0.1   &4&0.2&1& 560&0.2740&borderline\\

          2&  (L)BCD/LIRG& 0.75 & 0.4           &2  &  500   &   $\sim 40$  &3&0.5&1&1021&0.2060&accelerating \\

      3& Radio&1 &   0.5        &      2    &  1000   &   $\sim 200$  &3&0.7&1& 1208& 0.2946 &accelerating\\

\hline
\end{tabular}
 \medskip {\\
 Superwind hydrodynamical models.  Table headers: (a) steepness parameter, (b) concentration parameter, (c) radius, (d) dynamical mass, (e) star formation rate, (f) mass loading factor, (g) thermalization efficiency, (h) participation factor (i) effective terminal speed, (j) squared ratio of the escape velocity to the effective terminal speed, and (k) flow regime when the external halo is neglected.  }  \label{Table1}
\end{minipage}
\end{table*}

 \begin{table*} 
\begin{minipage}{175mm}
\caption{Reference hydrodynamical models. External halo parameters.}
\begin{tabular}{@{}llllllll}
\hline
Model  &   Type&$\alpha_1$  &$A_1$ & $r_{\rm D}$ & $M_{\rm D}$  &$M_{\rm t}$&Regime\\
&&&&(kpc)&($\times 10^{8}$ M$_{\odot}$)&  ($\times 10^{8}$ M$_{\odot}$)\\
&&$(\rm a)$&$(\rm b)$&$(\rm c)$&$(\rm d)$&$(\rm e)$&$(\rm f)$\\
 \hline

        1&  dE&0.75& 0.5 &1 & $M_{\rm DM}$ & $\sim 2.5 M_{\rm DM}$ & open-box enrichment\\
                      2&  LIRG& 0.75 &1& 5& $\sim 3.16 M_{\rm DM}$  & $\sim  4.76M_{\rm DM}$ & decelerating \\
      3& Radio&1 (3/4)&1&$R_{\rm hm}=2.5$&  $2M_{\rm DM}$ & $4M_{\rm MD}$& decelerating (open-box enrichment)\\

\hline
\end{tabular}
 \medskip {\\
External halo parameters for the models presented in Table \ref{Table1}.  Table headers: (a) steepness parameter, (b) concentration parameter, (c) 'disc' radius (d) 'disc' mass, (e) total mass, and (f) Regime.  In model 3, $R_{\rm hm}$  corresponds to the half-mass radius. }  \label{Table2}
\end{minipage}
\end{table*}

\section{Summary and conclusions} \label{con}

Here we have presented an analytic free superwind model that incorporates the effect of extended DM and BM haloes. We find that the gravitational field of the extended haloes associated to massive galaxies can drastically alter the free superwind velocity profile and enhance its observable properties. We also find that massive haloes  can also contribute to the inhibition of the superwind solution. 

In our model, the galaxy total mass (BM+DM), the mass contained within a  bulge or galaxy  nucleus (defined by the characteristic radius $r_{\rm sc}$), the mass up to the disc characteristic radius,  and the steepness and concentration of the external halo, are all correlated.  Since  the correlations  are nonlinear,  deviations from galaxy to galaxy are permitted, see Tables \ref{Table1} and \ref{Table2}.  Oppositely,  there is no correlation between  the above parameters and the concentration and steepness of the mass distribution for $r<r_{\rm sc}$. This is consistent with the results of the cosmological simulations carried out by  Abadi et al. (2010) Navarro et al (2010), in the sense that haloes are not strictly universal. This should be expected, as we based our model in the 'structural contraction' property  derived from their simulations. On the other hand, in their  extensive work, Salucci et al. (2007) found   that the previous parameters were correlated for spiral galaxies, and proposed universal rotation curves assuming  a  Burkert  (1995) profile  for the DM distribution. Our theoretical work diverges from theirs in that we considered additionally the mentioned inner concentration and steepness,  which traces starburst episodes. Such a consideration discards the possibility of universal halo profiles and rotation curves, since in general they will differ for $r<r_{\rm sc}$; however, the discrepancy will be reconciled at larger radii, and thus one could talk of an  'asymptotically universal' property, in the sense defined by Salucci et al. (2007).

From the theoretical point of view, the importance of the analytic solution here presented resides  in that it can be used to construct approximated superwind solutions  when departures from spherical symmetry are important. In such a case, not just the parameters of the galaxy  but also its morphology  will   determine  both the superwind hydrodynamics  and the fate of the ejected gas; specially, of metals. The connection of these features   with the observed dispersion of the  M-Z relationship will be discussed in a forthcoming work.

\end{document}